\documentclass[preprint,showpacs,aps,amsmath,amssymb,pre,preprintnumbers]{revtex4}
\newcommand{\BE}{\begin{equation}}
\newcommand{\EE}{\end{equation}}
\newcommand{\BA}{\begin{eqnarray}}
\newcommand{\EA}{\end{eqnarray}}

\usepackage{graphicx}
\usepackage{dcolumn}
\usepackage{bm}
\input epsf.tex

\begin{document}

\title{Acoustic imaging and collimating by slabs of sonic crystals made from arrays of rigid cylinders in air}

\author{Liang-Shan Chen} \affiliation{Department of
Physics, Fudan University, Shanghai, China}

\author{Chao-Hsien Kuo}\author{Zhen
Ye}\email{zhen@phy.ncu.edu.tw} \affiliation{Wave Phenomena
Laboratory, Department of Physics, National Central University,
Chungli, Taiwan, China}

\date{\today}

\begin{abstract}

We show some new properties of the acoustic propagation in
two-dimensional sonic crystals, formed by parallel rigid cylinders
placed in air. The transmission through slabs of sonic crystals
and the associated band structures are considered. It is shown
that within partial bandgaps, the waves tend to be collimated or
guided into the direction in which the propagation is allowed.
Such a feature also prevails in the situations in which deaf bands
appear. We show that within the partial bandgaps, a stable imaging
effect can be obtained for flat sonic crystal slabs, in analogy to
the cases with photonic crystals.

\end{abstract}

 \maketitle

The exciting phenomenon of band structures in sonic crystals
allows for many possible applications, such as sound shields and
acoustic filters \cite{sanchez1,sanchez2,rober,kuswa}. These
applications mostly rely on the existence of complete sonic
bandgaps in which acoustic waves are prohibited from transmission
in all directions. Later, it was experimentally demonstrated
\cite{cervera} that proper SCs can also make refractive devices
such as Fabry-Perot interferometers and acoustically converging
lenses. The mechanism for the acoustic refractive devices lies in
the linearly dispersive response of SCs in the low frequency
regimes. In this case, it has been shown that lenticular SCs not
only can behave as normal acoustic lenses, but also the focusing
effect can be well described by Lensmaker's formula\cite{Lens}.

Recently, a new property of photonic crystals has been discovered
and has attracted much attention\cite{Attention}. That is, it was
suggested\cite{Luo} that in certain frequency regimes, a flat slab
of PCs not only is able to make focused images, but also such an
imaging may be related to the superlensing phenomenon, a concept
conceived from the study of the so called left-handed
materials\cite{Pendry}. Further explorations\cite{LiZY,neg2}
indicate that good quality imaging can be indeed formed across
flat photonic crystal slabs, and the overall imaging properties
are mainly governed by the wave-guiding or self-collimation
effects, in the presence of partial bandgaps or anisotropic wave
scattering. The flat PCs slab imaging has subsequently been
observed in the experiment \cite{Nature,Note}. These discoveries
are expected to pave a new avenue for photonic crystal
applications in controlling optical flows\cite{neg5}.

Due to the similarities between acoustic and EM waves, acoustic
imaging by flat slabs of sonic crystals may be also possible. In
the present paper, we wish to explore this possibility, following
the work done on photonic crystals\cite{Luo,LiZY,neg2}. We will
study the imaging and wave-guiding effects of flat sonic crystal
slabs and find the underlying mechanism for the imaging. The
results reported here may lead to new applications of sonic
crystals, such as medical imaging, and manipulation of acoustic
flows.

The sonic crystal systems are from the existing
experiments\cite{cervera}; in this way, we expect that
experimental verifications may be readily performed. Consider $N$
straight rigid cylinders located in air. An acoustic line source
transmitting monochromatic waves is placed at a certain spatial
point. The scattered wave from each cylinder is a linear response
to the incident waves which are composed of the direct wave from
the source and the multiply scattered waves from other cylinders,
and subsequently contributes to the total waves. Such a scattering
process has been first formulated exactly by Twersky
\cite{twersky}. In this paper, we will use Twersky's formulation
to calculate the multiple scattering in the systems. For regular
arrays of cylinders, the band structures are computed by the
plane-wave method.

In the simulation, the rigid cylinders of radius $a$ are arranged
to form square arrays in air with lattice constant $d$. The
following parameters from the experiment\cite{cervera} are used in
the simulations: (1) the radius of the cylindrical rods is $1.5$
cm; (2) the lattice constant is $11.0$ cm; (3) The filling factor,
defined as the area occupation of the cylinders per unit area $f_s
= \pi(a/d)^2$, is equal to 0.05842; (4) the sound speed in air is
$v_a = 334$m/s.  Additionally, the acoustic transmission through
the cylinder arrays is normalized as $T = p/p_0$. The acoustic
intensity field is defined as $|P|^2$. We note here that the rigid
cylinders can be any solid rods. As shown in Ref.~\cite{sanchez2},
any material whose acoustic impedance with respect to the air
exceeds roughly 10 can be used as the composition material for the
rods.

\begin{figure}[hbt]
\begin{center}
\epsfxsize=5in \epsffile{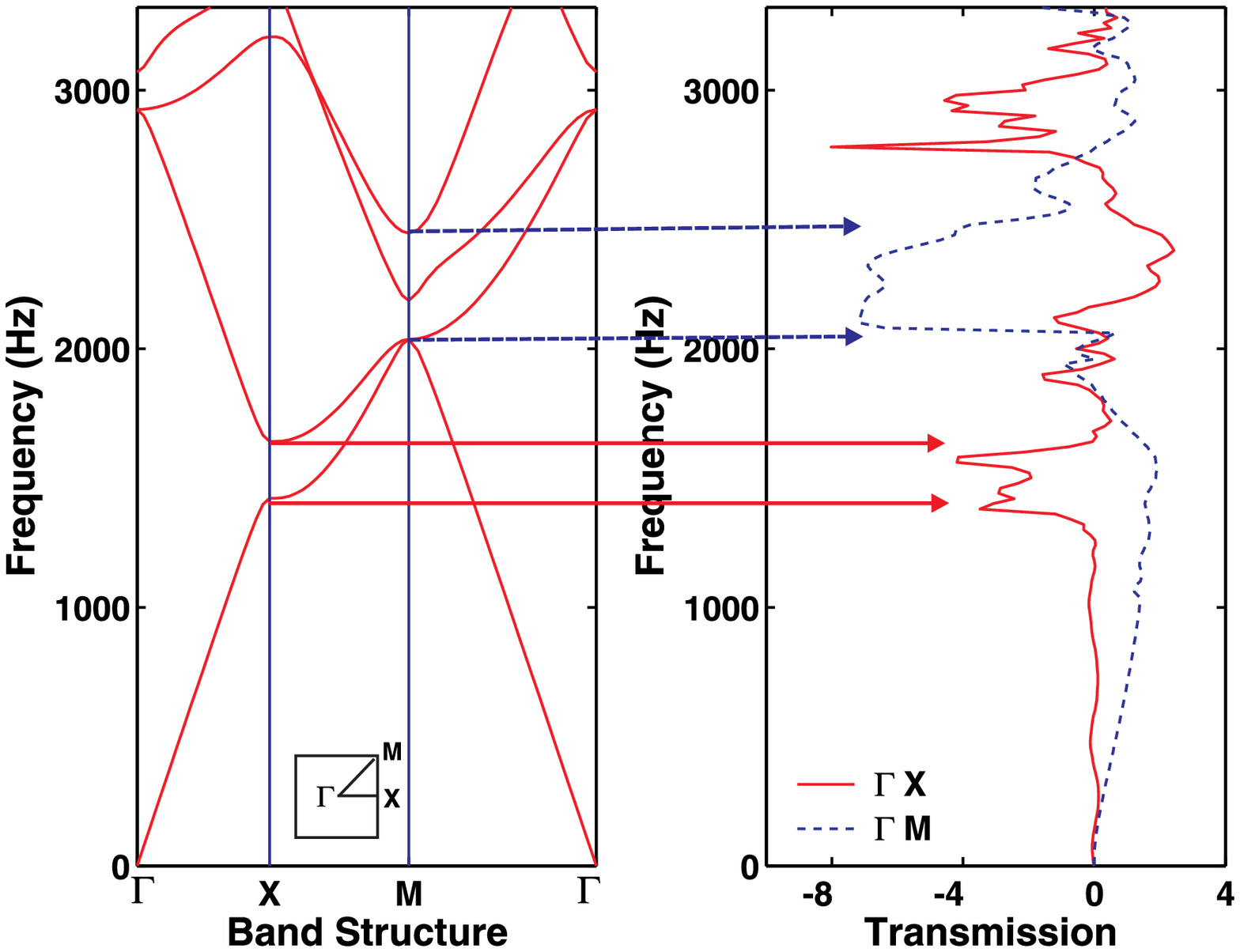} \caption{ \label{fig1}\small
Left panel: the band structure of a square lattice of rigid
cylinders in air. A partial gap is between the two horizontal
lines. Right panel: the normalized transmission versus frequency.
}
\end{center}
\end{figure}

Figure~\ref{fig1} presents the band structure and transmission
result for the square array of rigid cylinders. In the figure, the
curves on the left denote various dispersion bands when the wave
is propagating in different directions. The inserted box denotes
the first Brillouin zone. The results in Fig.~\ref{fig1} reproduce
nicely that in the experiment\cite{cervera}. Here is shown that
there is a partial bandgap ranging from 1.42 to 1.64 KHz, while a
deaf band region appears from 2.08 to 2.72 KHz. The partial gap
and the deaf band regimes are denoted by the horizontal lines
respectively. Within the partial gap, the waves are prohibited
from propagation along the $\Gamma X$ direction, i.~e. the [10]
direction. For frequencies inside the deaf band, the propagation
is inhibited along the $\Gamma M$ direction, i.~e. the [11]
direction.

We find that the flat slab imaging can be formed for frequencies
located within these regimes, in analogy with the photonic crystal
cases. For brevity, we choose two frequencies for simulation, that
is 1.50 KHz and 2.11 KHz which are within the partial gap and the
deaf band respectively. Hereafter we denote the two cases as Case
G and Case C respectively.

\begin{figure}[hbt]
\begin{center}
\epsfxsize=5in \epsffile{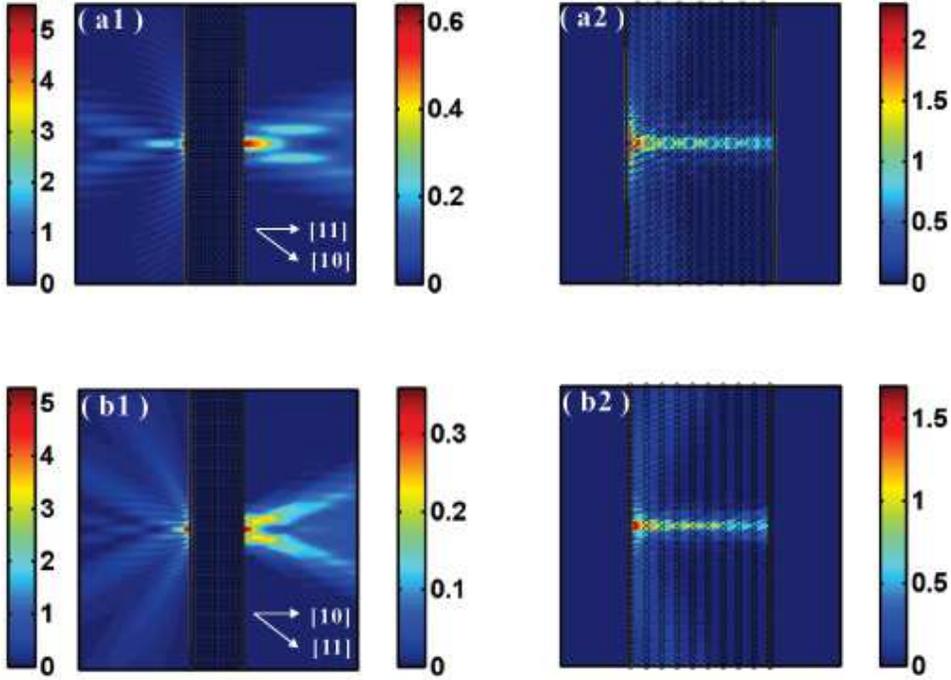} \caption{ \label{fig2}\small
The imaging fields for two slabs of sonic crystal structure for
two frequencies 1.50 KHz and 2.11 KHz, located within the partial
gap and the deaf band area respectively. (a1) and (b1) show the
intensity field outside the slab. (a2) and (b2) plot the intensity
fields for the areas inside. Here we see clearly that the waves
propagate mainly in a small tunnel inside the slab for both
frequencies, and make the focused images across the slab. }
\end{center}
\end{figure}

Figure~\ref{fig2} shows the imaging fields $|P|^2$ for two slabs
of sonic crystals at frequencies 1.5 KHz and 2.11 KHz, i.~e. Case
G and Case D, respectively. For Case G in (a1) and (a2), the array
measures as $7\sqrt{2}d\times 36\sqrt{2}d$, while the slab has the
size of $9d\times 49d$ for Case D in (b1) and (b2). The
orientations of the slabs are denoted in (a1) and (b1): the
transmission is along [11] in Case G and along [10] in Case D. The
acoustic source is placed at one lattice constant away from the
left sides of the slabs. The results for Case G and Case D are
similar. Here, we observe separately two well focused image points
on the right hand sides of the two slabs. When looking at the
intensity fields inside the two slabs ((a2) and (b2)), we find
that the transmission is mainly confined within a pipeline or
tunnel along $\Gamma M$ for Case G and along $\Gamma X$ in Case D,
and the waves travel to the other side of the slab, then release
into the free space. This type of self-collimating or guiding
behavior has been previously discussed for photonic
crystals\cite{LiZY,neg2}.

The imaging and wave-guiding results in Fig.~\ref{fig2} may be
understood in the framework of the band structure results. The
band structure from Fig.~\ref{fig1} clearly shows that in the
frequency ranges considered, there are a band gap along the
$\Gamma X$ direction, and deaf band region along the $\Gamma M$
directly. Therefore the waves are prohibited from propagation
along these two directions for Case G and Case D respectively. In
other directions such as $\Gamma M$ for Case G, however, there is
an allowed band to support the wave propagation. To be specific,
we consider Case G as the example. In the present setup, the
incident wave is set along the $\Gamma M$ direction, which makes
an angle of 45 degrees to the $\Gamma X$ direction. Since being
prohibited from propagating in the $\Gamma X$ direction, i.~e. 45
degrees from the straight horizontal direction, waves tend to move
forward along the allowable $\Gamma M$ direction. The frequency
band in the $\Gamma M$ direction thus provides a propagating
avenue for the waves to go over to the other side of the slab,
forming the images. As to be published elsewhere, we note that the
results from Fig.~\ref{fig2} may not be explained by the
equal-frequency method proposed in Ref.~\cite{Luo}.

\begin{figure}[hbt]
\begin{center}
\epsfxsize=5in \epsffile{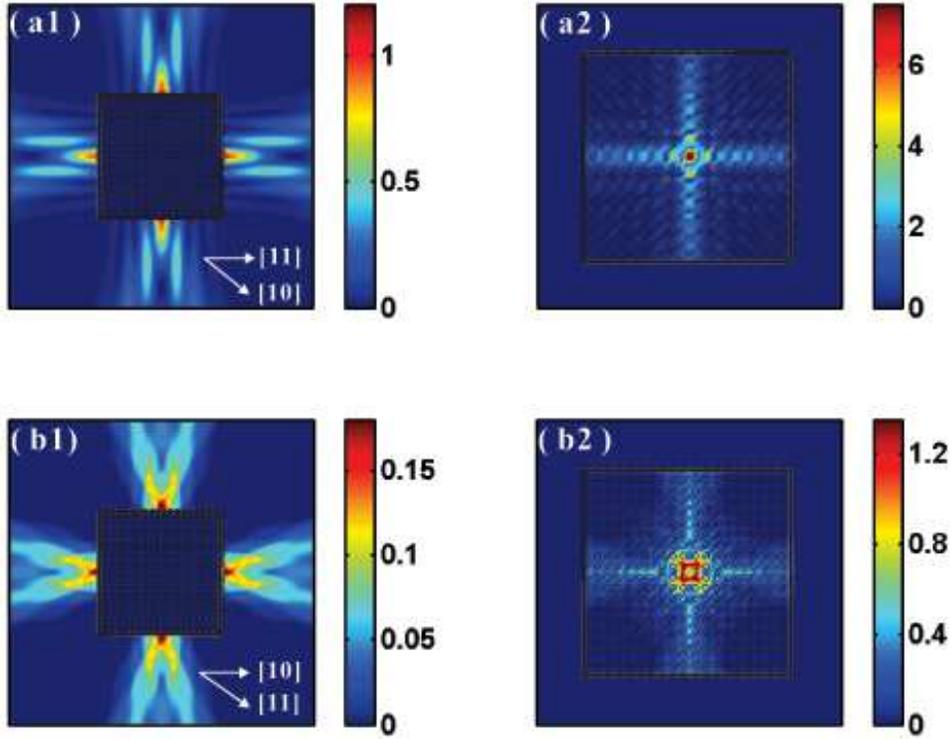} \caption{ \label{fig3}\small
The imaging fields for a transmitting source located inside two
square arrays of cylinders. In (a1) and (a2), the square measures
$14\sqrt{2}\times 14\sqrt{2}$, while the square array has the size
of $19\times 19$ in (b1) and (b2). All other parameters are the
same as in Fig.~\ref{fig2}.  (a1) and (b1) show the intensity
field outside the slab. (a2) and (b2) plot the intensity fields
for the areas inside the slabs.}\end{center}
\end{figure}

To further support our observation, we place a transmitting source
{\it inside} two arrays of cylinders. The overall shape of the
array is square. Fig.~\ref{fig3} presents the simulation results.
The cylinders are arranged to form a square arrays whose sides
measure respectively 14$\sqrt{2}d$ for Case G and $19d$ for Case
D. Same as in Fig.~\ref{fig2}, the orientations of the slabs are
denoted in (a1) and (b1). The intensities for waves outside or
inside the slabs are shown in the (a1), (b1) and (a2), (b2) for
the two cases. As expected, the focused images are evident in four
allowed directions for each case: in the [1,1], [-1,1], [-1,-1],
and [1,-1] directions for Case G, and in the [1,0], [0,1], [-1,0],
and [0,-1] for Case D. The field images inside the arrays clearly
show the travelling path of the waves along these directions,
depicted by (a2) and (b2).

\begin{figure}[hbt]
\begin{center}
\epsfxsize=5in \epsffile{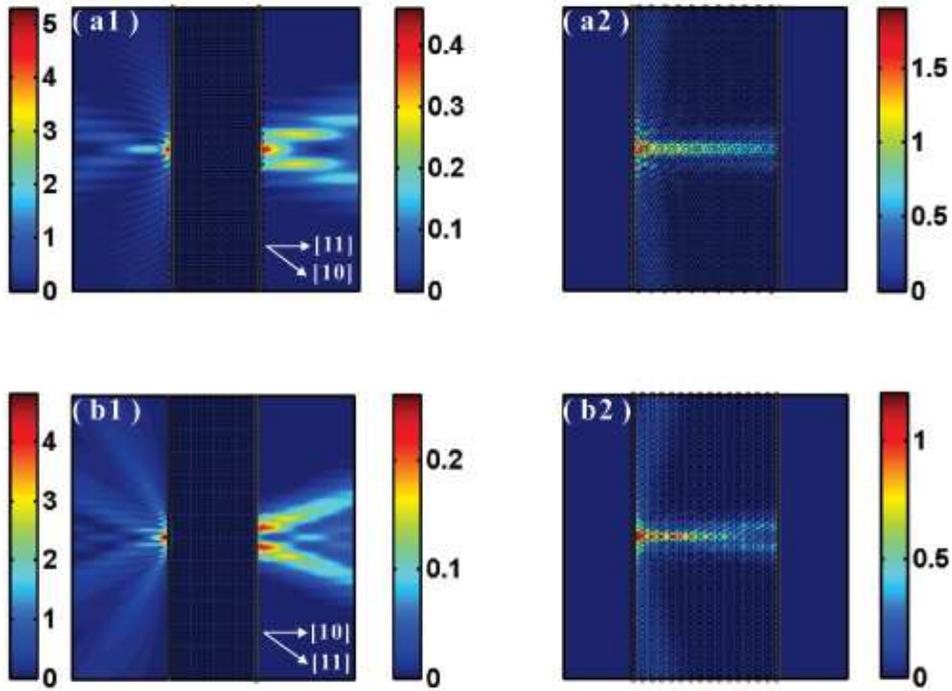} \caption{ \label{fig4}\small
The imaging fields for two slabs of sonic crystal structure,
similar to Fig.~\ref{fig2} but with larger sample lengths. }
\end{center}
\end{figure}

\begin{figure}[hbt]
\begin{center}
\epsfxsize=5in \epsffile{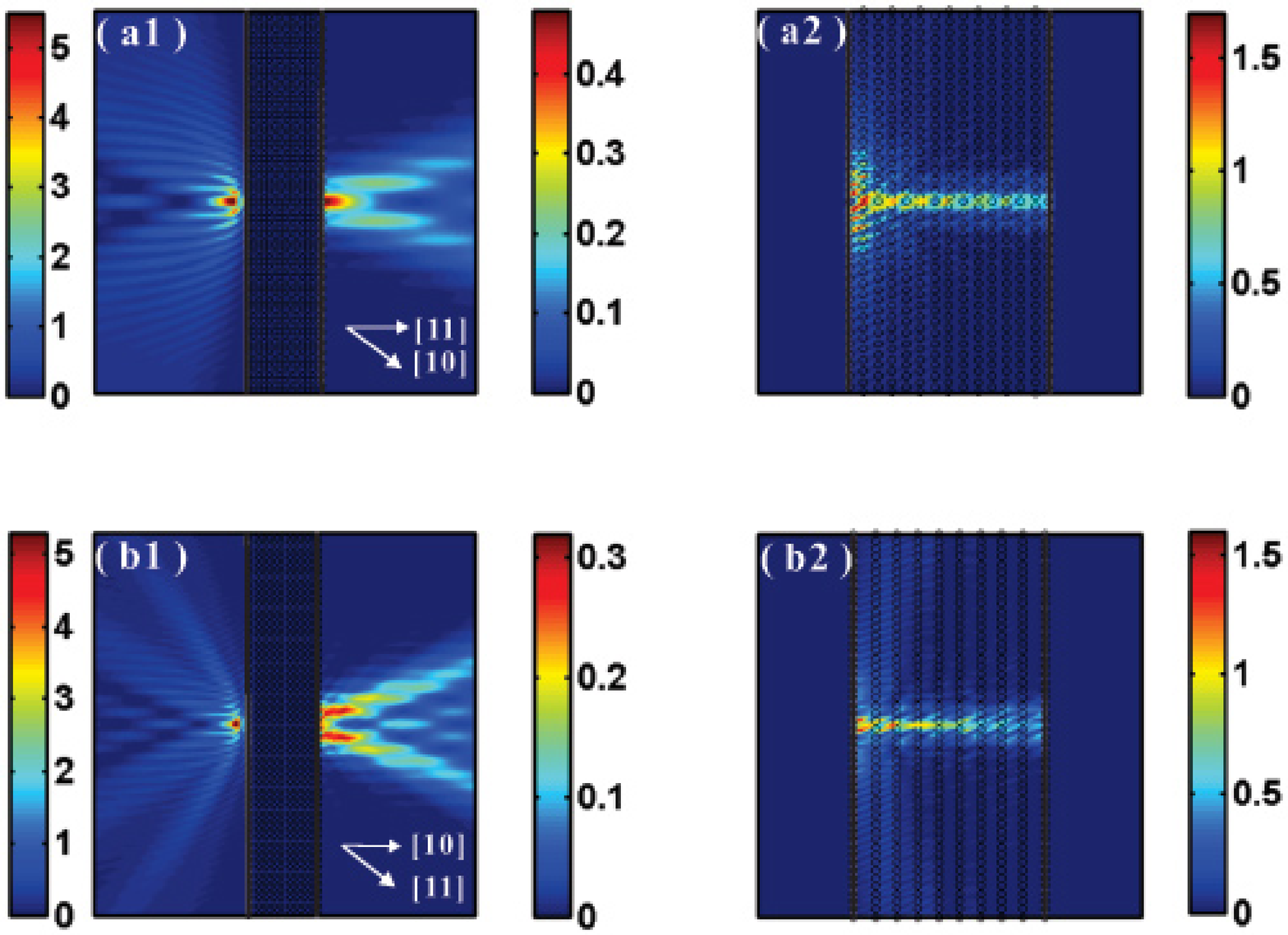} \caption{ \label{fig5}\small
The imaging fields for two slabs of sonic crystal structure,
similar to Fig.~\ref{fig2} except that the source has been moved
one more lattice constant away from the slabs. }
\end{center}
\end{figure}

The above flat slab imaging and guiding results are qualitatively
similar for both partial gaps and the deaf bands. Further
simulations indicate that the imaging effect for partial gaps is
insensitive to either source location nor the slab horizontal
size, revealing a stable imaging. For the deaf band situation,
however, the cross slab imaging can depend on the the location or
the slab size. In Figs.~\ref{fig4} and \ref{fig5} we show the
imaging effects for two variations of the results in
Fig.~\ref{fig2}. In Fig.~\ref{fig4}, the slab sizes have been
increases to be $11\sqrt{2}d\times 36\sqrt{2}d$ for Case G and
$15d\times 49d$ for Case D. In Fig.~\ref{fig5}, the source is
moved one more lattice constant away from the slab. Here, we
observe that the imaging patterns for Case G do not change: there
is still a single well formed imaging point with nearly the same
distance to the slab. In Case G, however, the single imaging point
is split into two focused points. The feature of guided travelling
paths inside the slabs, however, remains unchanged for both cases.

In summary, following the recent development in the photonic
crystal research, we propose some new properties of sonic
crystals. That is, a flat slab of sonic crystals can make well
focused images. This peculiar effect can be caused by either the
partial bandgaps or the deaf bands. It is also found that the
partial bandgaps or the deaf bands can guide wave propagation into
passing band directions.

The work received support from National Science Council. We
greatly appreciate the correspondence with M. Nieto-Vesperinas, J.
Sanchez-Dehesa, J. Zi, and J. Pendry.

\newpage

\newpage

\newpage

\newpage

\newpage

\begin{thebibliography}{99}
\bibitem{Note} We note that the interpretation of the slab imaging effects in
\cite{Nature} is based upon the theoretical analysis in
Ref.~\cite{Luo}. The latter has been questioned in
Refs.~\cite{LiZY,neg2}.
\end{thebibliography}
\end{document}